\documentclass{optica-article}

\journal{opticajournal} 

\articletype{Research Article}

\usepackage{lineno}
\usepackage{seqsplit}

\begin{document}

\title{
Enhanced Secrecy in Optical Communication using Speckle from Multiple Scattering Layers
}

\author{Alfredo Rates,\authormark{1,*} Joris Vrehen,\authormark{2} Bert Mulder,\authormark{1} Wilbert L. IJzerman,\authormark{2,3} and Willem L. Vos\authormark{1}}

\address{\authormark{1}Complex Photonic Systems (COPS), MESA+ Institute for Nanotechnology, 
University of Twente, P.O. Box 217, 7500 AE Enschede, The Netherlands\\
\authormark{2}Signify NV, High Tech Campus 7, 5656 AE Eindhoven, The Netherlands\\
\authormark{3}Department of Mathematics and Computer Science, Eindhoven University of Technology (TU/e), 5600 MB Eindhoven, The Netherlands}

\email{\authormark{*}a.ratessoriano@utwente.nl}

\begin{abstract*} 
We study the secrecy of an optical communication system with two scattering layers, to hide both the sender and receiver, by measuring the correlation of the intermediate speckle generated between the two layers. 
The binary message is modulated as spatially shaped wavefronts, and the high number of transmission modes of the scattering layers allows for \textit{many} uncorrelated incident wavefronts to send the \textit{same} message, making it difficult for an attacker to intercept or decode the message and thus increasing secrecy. 
We collect 50,000 intermediate speckle patterns and analyze their correlation distribution using Kolmogorov-Smirnov (K-S) test. 
We search for further correlations using the K-Means and Hierarchical unsupervised classification algorithms. 
We find no correlation between the intermediate speckle and the message, suggesting a person-in-the-middle attack is not possible. 
This method is compatible with any digital encryption method and is applicable for codifications in optical wireless communication (OWC).
\end{abstract*}

\section{Introduction}\label{sec:intro}

\begin{figure}[tbp]
    \centering
    \includegraphics[width=0.5\columnwidth]{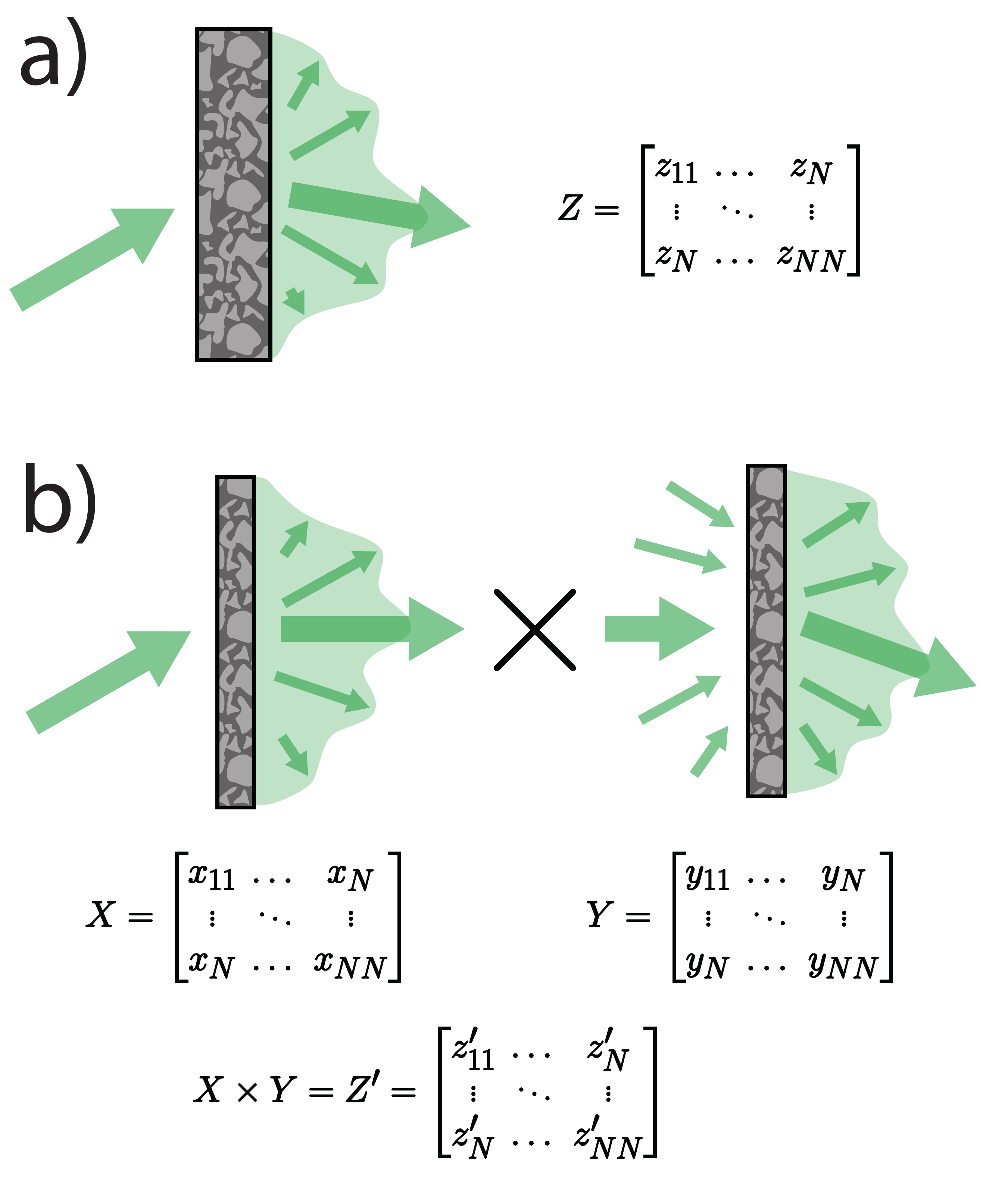}
    \caption{Schematic of the light scattering systems studied here.
    (a) A single incident light beam is sent through $N=1$ scattering slab, yielding a complex speckle pattern as output. 
    The speckle pattern transmitted through the slab is described by a transmission matrix $Z$. 
    (b) A single incident light beam is sent through $N=2$ scattering slabs. 
    The speckle pattern emanating from the first slab is the intermediate pattern (described by transmission matrix $X$) that is sent onto the second slab (with transmission matrix $Y$). 
    The final output is a new speckle pattern described by a transmission matrix $Z^\prime$ with $Z^\prime = X \times Y$.} 
    \label{fig:multiLayer}
\end{figure}

Scattering of light occurs in any 3D complex opaque material such as paint, foam, or biological tissue, independent of its shape: whether slab, fiber, or free-form~\cite{Ishimaru1978book, vanRossum1999RevModPhys, Akkermans2007book, Carminati2021book}.
When light travels through a complex material, it takes many different paths or channels inside the medium, whereby the light performs a random walk with a typical step size called the mean free path~\cite{vanRossum1999RevModPhys}. 
Along these contorted paths, the light waves pick up a random distribution of phase changes.
Therefore, it is intuitively reasonable that a random interference pattern appears in a target plane, called a speckle pattern, that consists of a random arrangement of bright and dark areas.
While a random speckle interference also arises due to scattering from a rough surface~\cite{Goodman2007book}, the speckle due to light scattering in a 3D complex material has several additional physical properties, notably, a number of intricate correlation functions, see Refs.~\cite{vanRossum1999RevModPhys, Akkermans2007book}. 
Since the arrangement of speckle spots in an observation plane is exceedingly difficult to predict due to the huge number of degrees of freedom in a complex material, the complex light scattering in such materials offers an attractive opportunity to encode information; indeed, complex materials play a central role in optical physically unclonable functions (PUF) and are employed in optical cryptography~\cite{Pappu2002science, Ruhrmair2010Book, Goorden2014optica, Dolev2016CompComm, Tehranipoor2023Book}. 

Although a speckle pattern is random, the intensities of many single speckle spots have a well-defined exponential distribution, also known as Rayleigh distribution, that is characterized by the average intensity. 
The transport of scattered light through a complex material is successfully described by random matrix theory from mesoscopic physics, that invokes a large transmission matrix with many complex-valued elements~\cite{Beenakker1997RevModPhys, Kohlgraf2008OptExp, Popoff2010PRL, Akbulut2016PRA}, illustrated in Fig.~\ref{fig:multiLayer}a). 
Knowledge of the transmission matrix has been successfully extracted and even applied to imaging and transmitting encoded data~\cite{Popoff2010PRL, Popoff2010NatComm}. 

Statistics of the transmission matrix reveal the existence of ``closed channels'', with zero transmission, and ``open channels'', with almost perfect transmission~\cite{Dorokhov1984SSC, Mello1988AnnPhys, Imry1986EPL, Pendry1990}. 
So even a thick complex scattering medium is effectively transparent thanks to these states~\cite{Vellekoop2008PRL}. 
Modulation techniques such as wavefront shaping (WFS) and mutual scattering profit from the high-transmission states to control the transmission through the scattering medium, thereby modulating the intensity distribution of speckle spots~\cite{Vellekoop2007OptLett, Vellekoop2008PRL, Lagendijk2020EPL, Rates2021PRA}. 
This can be used, notably, to concentrate the intensity at a specific focal point, transmit information, or change the overall transparency of the scattering medium.

When the properties of the incident light are changed (\textit{e.g.}, positioned, tilted, shaped, or frequency shifted), the light is scrambled in a different way resulting in a new speckle pattern in the observation plane. 
When the range of perturbations of the incident light is moderate, known as the memory range, the new speckle pattern has remarkable non-zero correlations with the original speckle pattern~\cite{Freund1988PRL}.
This effect is typically characterized by intensity-intensity correlations between pairs of different positions in \textit{one observation plane}, known as C1 correlations~\cite{Shapiro1986PRL, vanRossum1999RevModPhys}. 
Such speckle and intensity correlations find many applications, ranging from imaging through an opaque screen~\cite{Bertolotti2012Nature, Yilmaz2015Optica} to transmitting images and information through opaque media~\cite{Cao2022ole}.

In this work, we study speckle correlations in a different setting, where we use two layers of scattering media and observe the \textit{intermediate} speckle pattern in between the scattering layers. 
Instead of measuring changes of a speckle pattern in a target plane - as is described above - we collect two speckle patterns in \textit{two different observation planes}, namely one in between the two scattering layers and the other after passing through both scattering layers, illustrated in Fig.~\ref{fig:multiLayer}b). 
For applications to optical wireless communication, we study if a binary message sent through these scattering layers can be extracted or not by only observing the intermediate speckle pattern.
The presence of the first scattering layer is essential to spatially scrambling the incident wavefront, effectively ``hidding''  the sending modulator for direct inspection and subsequent decoding. 
The question we address in this work is whether different incident wavefronts are correlated with each other, knowing the message they result into. 
In other words, if the two transmitted wavefronts correlate, does that mean the two intermediate wavefront are also correlated?

We synthesize $N_{\rm W}=50,000$ different incident wavefronts and collect the resulting intermediate and final speckle patterns. 
The message is encoded as an intensity distribution in the final speckle pattern, where each speckle pattern is assigned to either a binary 0 or a binary 1. 
Thus, we have thousands of available wavefronts to send the same value. 
We measure the correlation between intermediate speckle patterns, to study if the intermediate speckle patterns resulting in the same message encoded in the final speckle pattern are somehow correlated. 

%
%
\section{Working principle and experimental methods}\label{sec:exp}
%

In light scattering theory, and in general in mesoscopic physics, a single thick slab of scattering material with thickness $L$ is equivalent to $N$ thin slabs with average thickness $(L/N)$ since the total thickness is the same: $N (L/N) = L$. 
Here, we study $N=2$ as illustrated in Fig.~\ref{fig:multiLayer}. 
A scattering material is represented as a random transmission matrix that couples the modes of the incident light to the ones of the outgoing light, and the random components of the transmission matrix represent the scattering events inside the material~\cite{Beenakker1997RevModPhys, vanRossum1999RevModPhys, Miller2019AdvOptPhot}. 
Studying $N=2$ scattering slabs is represented as the multiplication of two of these random matrices, which because it is a linear system, results in yet another random transmission matrix. 
The reasoning above implies that techniques such as wavefront shaping or mutual scattering are also relevant and valid when multiple slabs are used, independent of the distance between the slabs. 
Here, we study if there are correlations between the intermediate speckle pattern that we observe between the two scattering slabs on the one hand and the output speckle pattern on the other hand. 

\begin{figure}[tbp]
    \centering
    \includegraphics[width=\columnwidth]{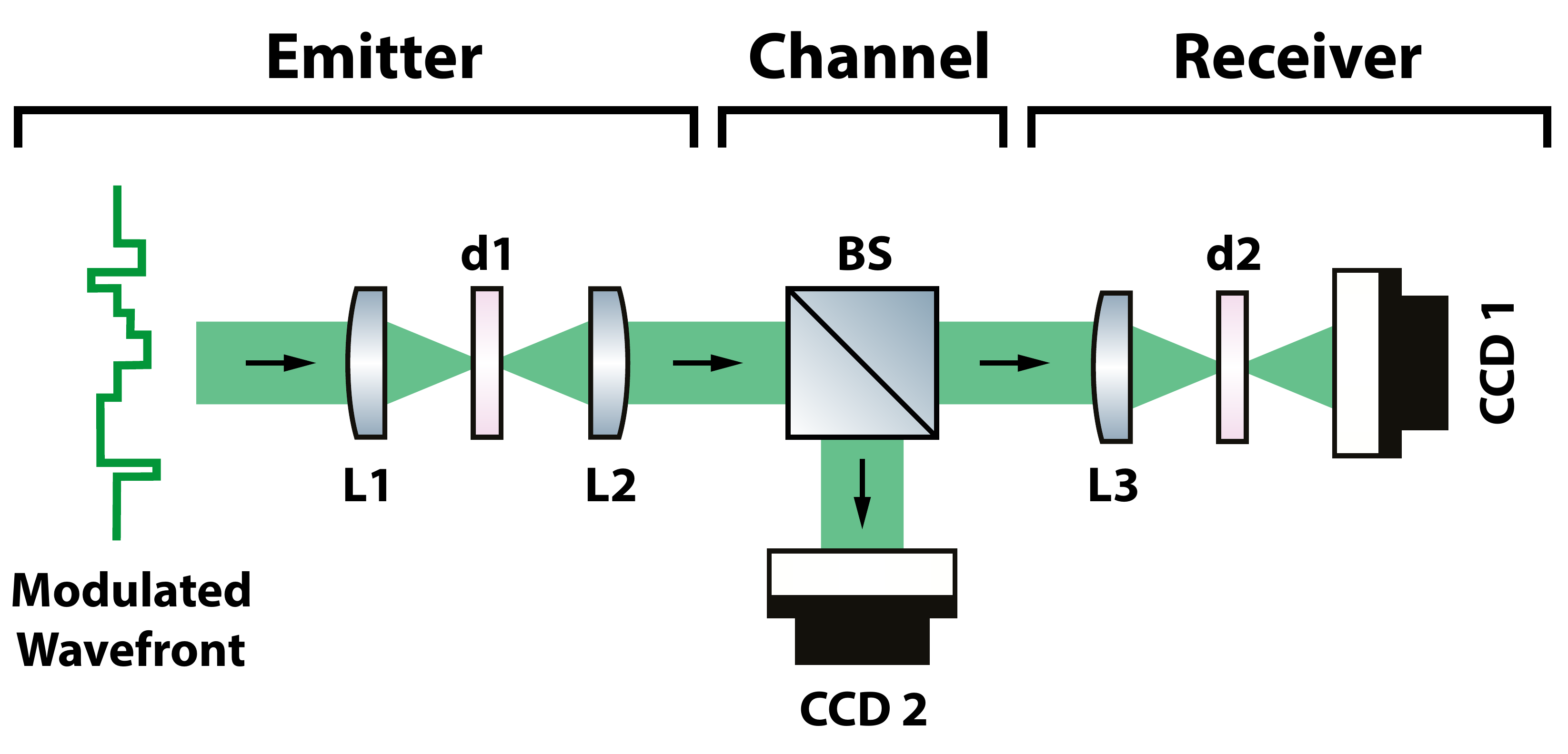}
    \caption{Diagram of the experimental setup. 
    The incident wavefront is spatially phase-modulated using a Digital Mirror Device (DMD, not shown)~\cite{akbulut2011optexp}. 
    The diffusers $d_1$ and $d_2$ are scattering materials forming speckle patterns. 
    Camera CCD1 records the final speckle pattern after both $d_1$ and $d_2$ and CCD2 records the intermediate pattern after $d_1$ only.
    (L: Lens, BS: Beamsplitter).}
    \label{fig:setup}
\end{figure}

To study the speckle correlation, we used the experimental setup shown in Fig.~\ref{fig:setup}. 
The initial light source is a frequency-doubled continuous wave green ($\lambda=532$\ nm) Nd:YAG$^3+$ laser (Coherent Compass 315M-100, 100mW). 
The signal is encoded as a phase-modulated wavefront using a digital mirror device (DMD, Vialux VX4100), in the same way as usually done in Wavefront Shaping experiments~\cite{akbulut2011optexp}. 
The DMD applies a binary, ON-OFF modulation to the wavefront, which we transform into a phase modulation using the Lee Holography technique~\cite{Lee1978ProgrOpt}. 
Using phase modulation rather than amplitude-only modulation we achieve greater control over the scattering events.~\cite{Lee1978ProgrOpt, Conkey2012OptExp} 
The modulated wavefront is focused into a diffuser (Ground Glass Diffuser 220 Grit, Thorlabs) using lens L1 ($f=75$ mm), and collimated using a microscope objective (NA=0.3, Nedoptifia Zeist).
This diffuser has a large diffusing pattern and high transmission, and it is commonly used in scattering experiments~\cite{Bertolotti2012Nature, Edrei2016SciRep}.
To measure the light speckle between the diffusers, a beamsplitter is used to pick up half of the signal, which is collected by a charge-coupled device (CCD) camera (Guppy F-146B), to detect the intermediate speckle pattern.
The other half is focused on a second diffuser similar to the first one using lens L3 ($f=50$ mm), and collected by a second CCD camera (Stingray F-125), which we call the receiver. 
The CCD camera is placed in the far field, at a distance of $30$ mm from the second diffuser.

Light coherence is a key factor in our experiment.
We need spatial coherence in order to obtain and measure the speckle pattern. 
In our experiment, we use a well-defined laser beam as a source, meaning we have a high temporal and spatial coherence. 
 The laser Coherent Compass 315M-100 is estimated to have a line width of around 10MHz, and a coherent length on the order of tens of meters~\cite{laserBlog}, by far sufficient for our purposes. 
Our experiment has a path length on the order of one meter, thus ensuring the wavefront is spatially coherent.

%
%
\section{System characterization}\label{sec:charact}
%

The goal of our experiment is to send a signal through the two scattering slabs, and study if there is any correlation between the intermediate speckle pattern and the resulting pattern. 
As a starting point, we aim to obtain a binary signal, \textit{i.e.}, only two levels: state 0 and state 1. 
Obviously, the modulated incident wavefront has many more degrees of freedom than these two states. 
For this reason, there exist multiple incident wavefronts, and thus multiple intermediate speckle patterns, that can result in the same signal. 
We use this property to see if different intermediate speckle patterns need to have some degree of correlation in order to result in the same final state.

\begin{figure}[tbp]
    \centering
    \includegraphics[width=\columnwidth]{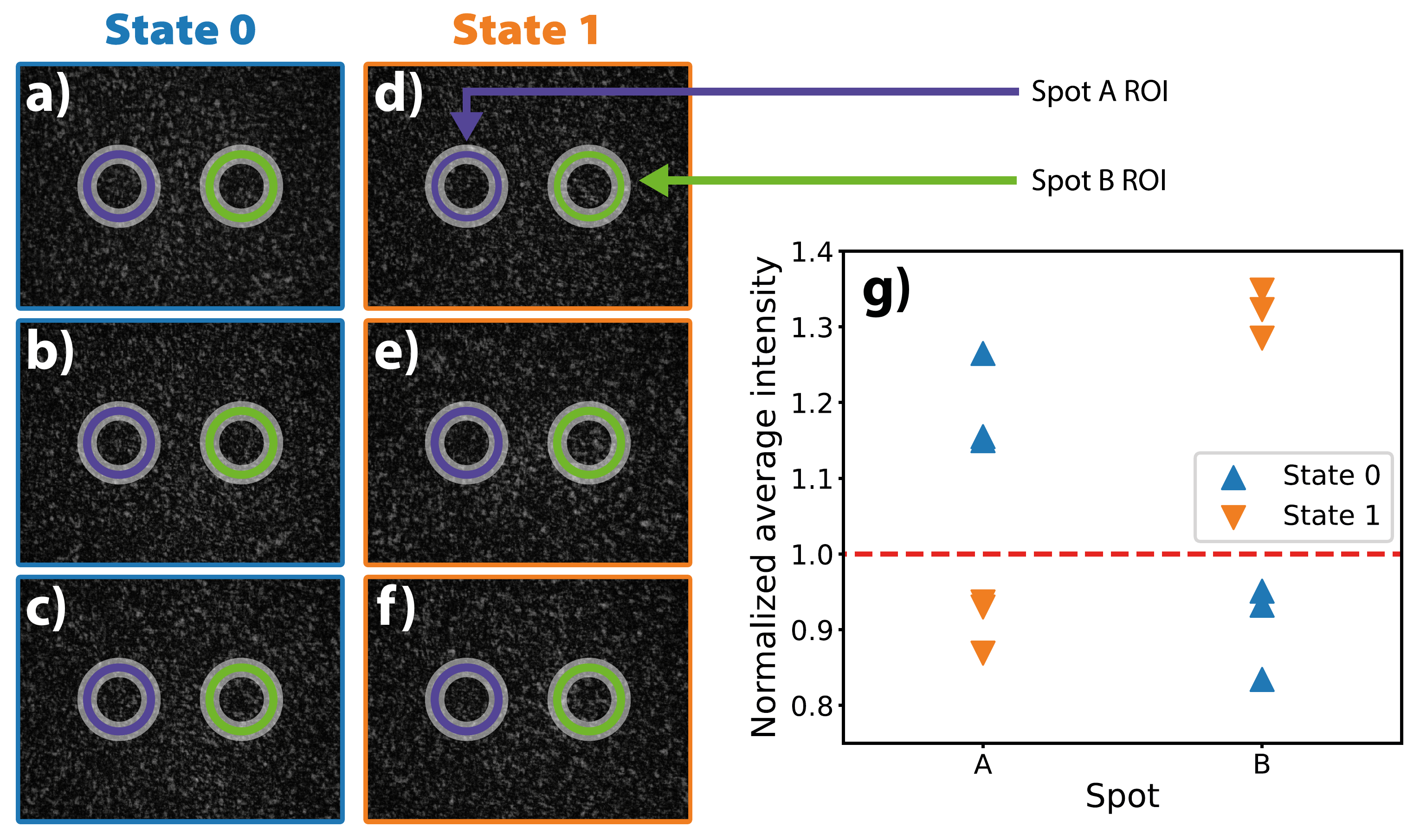}
    \caption{Examples of observed speckle patterns. 
    (a-c) Speckle patterns selected as state 0. 
    (d-f) Speckle patterns selected as state 1. 
    The colored circles in each panel show the region of interest (ROI) of spot A (purple) and spot B (green). 
    (g) Average intensity normalized by the average intensity of the background of spot A and spot B for each speckle pattern selected as state 0 or state 1.
    }
    \label{fig:speckles}
\end{figure}


To be robust against environmental noise and to have a large dynamic range, we consider in the receiver plane only two regions of interest (ROI), called spot A and spot B, as shown in Fig.~\ref{fig:speckles}. 
From the intensity distribution of the speckle at the receiver, we estimate this area to be $\times 15$ larger than the size of a single speckle. 
Only when the local average intensity of spot A is high and the local intensity of spot B is low, a state 0 is received. 
Conversely, if the local intensity of spot A is low and the local intensity of spot B is high, state 1 is received, see Fig.~\ref{fig:speckles}g). 

\begin{figure}[tbp]
    \centering
    \includegraphics[width=\columnwidth]{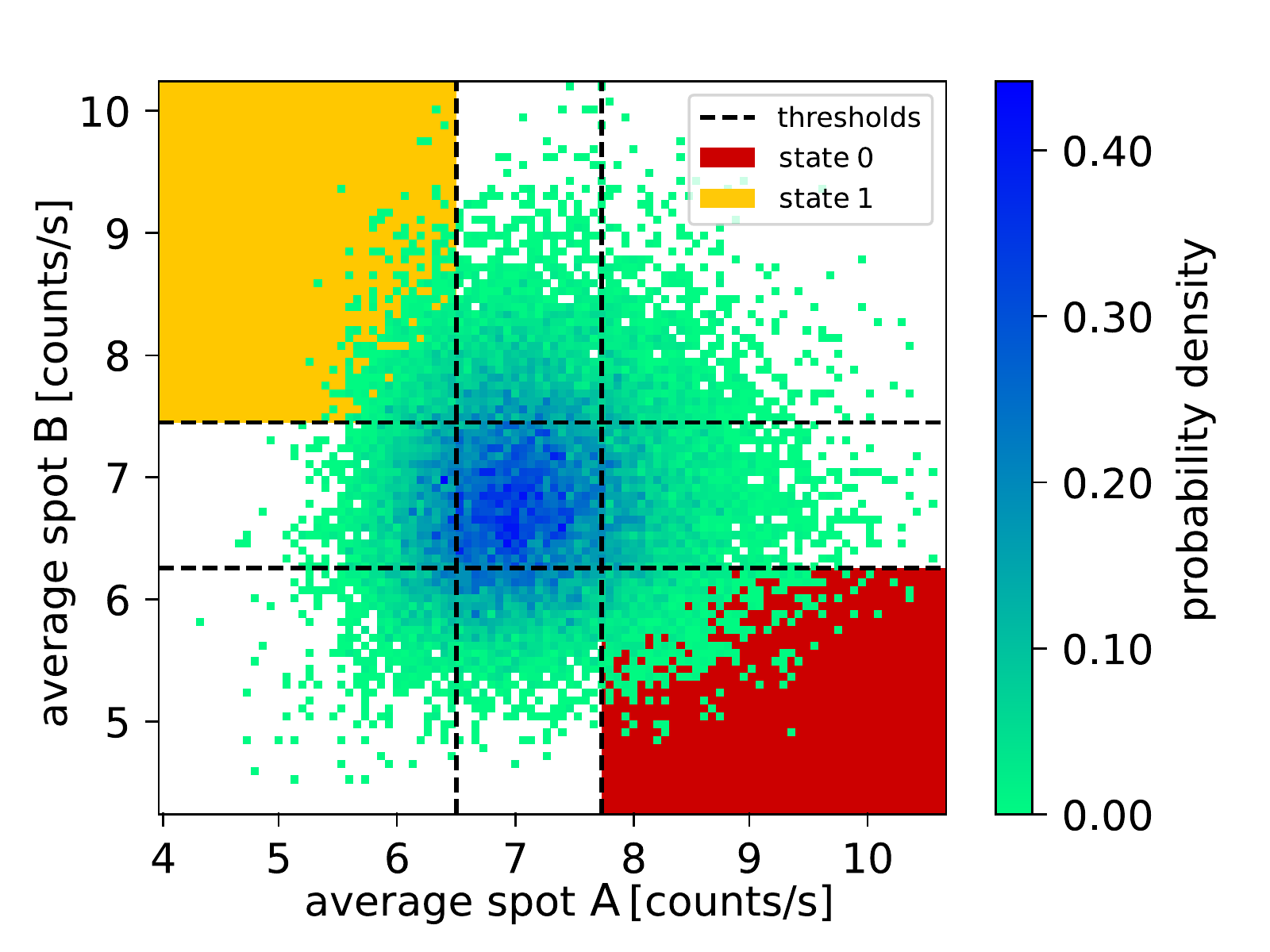}
    \caption{Joint distribution of average intensities at spot A and at spot B at the receiver for different incident wavefronts (green symbols). 
    The horizontal and vertical dashed lines are the intensity thresholds that define the states 0 or 1. 
    Here, the red region indicates the accepted state 0 with high intensity in spot A and low intensity in spot B. 
    The blue region indicates the accepted state 1 with low intensity in spot A and high intensity in spot B. 
    }
    \label{fig:intensities}
\end{figure}

To select the speckle patterns corresponding to state 0 or state 1, we send a total of $N_{\rm W}=50,000$ randomized wavefronts. 
For each wavefront, the speckle pattern at CCD2 is collected along with the intensities at spot A and spot B. 
From the definition of each state, we define arbitrary intensity thresholds based on the joint intensity distribution of spot A and spot B, shown in Fig.~\ref{fig:intensities}, where the thresholds are marked with dashed lines. 
For both spot A and spot B, we set the thresholds at 20\% and 80\%. 
This means that a speckle pattern is classified as state 0 only if the intensity at spot A is higher than 80\% of the distribution, and the intensity at spot B is lower than 20\% of the distribution. 
The classification of wavefronts as state 1 follows the same principle. 

The red and blue regions in Fig.~\ref{fig:intensities} highlight which speckle patterns are accepted as state 0 or state 1, respectively. 
With the given thresholds, there is a total of $N_{\rm W,0}=1839$ available wavefronts (3.7\% of the total) to get a state 0, and $N_{\rm W,1}=1774$ (3.5\% of the total) to get a state 1.
We modulate the wavefront using a grid of $15\times15$ segments, controlling the phase of each segment. 
In our current realization, we modulate the phase from $0$ to $2\pi$ in 16 steps. 
That means that the maximum number of different wavefronts that we generate is equal to $N'_{\rm W}=16^{225}$, which is on the order of $\mathcal{O}(10^{307})$. 
The large complexity of the scattering material is such, that small changes in the phase of a single segment at the modulator produce large changes in the intensity distribution at the receiver.
This means that the number of available wavefronts can be made arbitrarily large, at the expense of longer measurements and digital memory.

%
%
\section{Correlation distribution of intermediate speckles}\label{sec:correlation}
%

Every wavefront used to send a message also generates an intermediate speckle pattern between the slabs. 
When sending two messages separately, we generate two intermediate speckles and we calculate the correlation between these two intermediate speckles. 
We are interested in comparing three cases: the correlation between two speckle patterns from state 0 (0-0 correlation), the correlation between two speckle patterns from state 1 (1-1 correlation), and the correlation between a speckle pattern from state 0 and another from state 1 (0-1 correlation). 

\begin{table}[tbp]
\centering
\caption{Example of pairs of intermediate speckle patterns used to calculate the Pearson correlation distribution for each relevant case.
The left column indicates the case, while the right column shows each pair used for that case.
The letters between parentheses refer to the sub-figure indexes of the speckle patterns shown in Fig.~\ref{fig:speckles}.}
\label{tab:exCorr}
\begin{tabular}{l|l}
\textbf{Case}   & \textbf{Pairs of speckle patterns}                                                                        \\ \hline
0-0 correlation & (a,b), (a,c), (b,c)                                                                                       \\ \hline
1-1 correlation & (d,e), (d,f), (e,f)                                                                                       \\ \hline
0-1 correlation & \begin{tabular}[c]{@{}l@{}}(a,d), (a,e), (a,f),\\ (b,d), (b,e), (b,f),\\ (c,d), (c,e), (c,f)\end{tabular}
\end{tabular}
\end{table}

We illustrate the combinations of speckles in Table~\ref{tab:exCorr} using the panels indexes of Fig.~\ref{fig:speckles}.
To characterize the correlations between speckle patterns, we use the Pearson correlation coefficient $C_{\rm P}$ from the Python library SciKit~\cite{Pedregosa2011JMLR}, calculated as follows:
\begin{equation}\label{eq:pearson}
    C_{\rm P} \coloneqq \frac{\sum_{\rm i,j}(x_{\rm i,j}-\overline{x})(y_{\rm i,j}-\overline{y})}{\sqrt{(\sum_{\rm i,j}(x_{\rm i,j}-\overline{x})^2(y_{\rm i,j}-\overline{y})^2}},
\end{equation}
with $x_{\rm i,j}$ the value of pixel $(i,j)$ of the first speckle image, $y_{\rm i,j}$ the value of the pixel $(i,j)$ of the second speckle, and $\overline{x}$,\ $\overline{y}$ the average pixel values of the first and second speckle, respectively.

We calculate the correlation $C_{\rm P}$ between two intermediate speckles $s_{\rm int}$, \textit{e.g.}, $C_{\rm P}(s_{\rm int, 1},s_{\rm int, 2})$. 
Whether a message is classified as a state 0 or a state 1 does not depend on the intermediate speckle $s_{\rm int}$ but on the final speckle $s_{\rm f}$. 
The final speckle is expressed as $s_{\rm f}=Ys_{\rm int}$, with the transmission matrix of the second slab as $Y$ (see Fig.~\ref{fig:multiLayer}). 
We thus choose states 0 and 1 as the set of final speckles $s_{\rm f}$ that have a high correlation with the target patterns $s^{\star}_{0}$ and $s^{\star}_{1}$, respectively. 
Following what we showed in section~\ref{sec:charact}, the target pattern $s^{\star}_{0}$ has a high intensity in spot A, zero intensity in spot B, and average intensity elsewhere .
Conversely for $s^{\star}_{1}$. 
Thus, an intermediate speckle results in a state 0 if $C_{\rm P}(Ys_{\rm int},s^{\star}_{0}) \approx 1$, and in a state 1 if $C_{\rm P}(Ys_{\rm int},s^{\star}_{1}) \approx 1$.

When measuring the correlations for the cases 0-0, 1-1, and 0-1, we investigate if the intermediate speckle pattern correlate, given that their resulting speckle correlate. 
If any correlation is needed to result in the same state, we expect the correlation between two patterns from the same group to be larger than between two patterns from different groups, \textit{e.g.}, we expect the 0-0 and 1-1 correlations to be larger than the 0-1 correlation.
For perspective, previous studies of speckle correlation, such as the memory effect, focus on the auto correlation of the resulting speckle $C_{\rm P}
(s_{\rm f, 1},s_{\rm f, 2})$ or in the correlation between the incoming and outgoing pattern $C_{\rm P}(s_{\rm int, 1},s_{\rm f, 1})$. 

\begin{figure}[tbp]
    \centering
    \includegraphics[width=0.5\columnwidth]{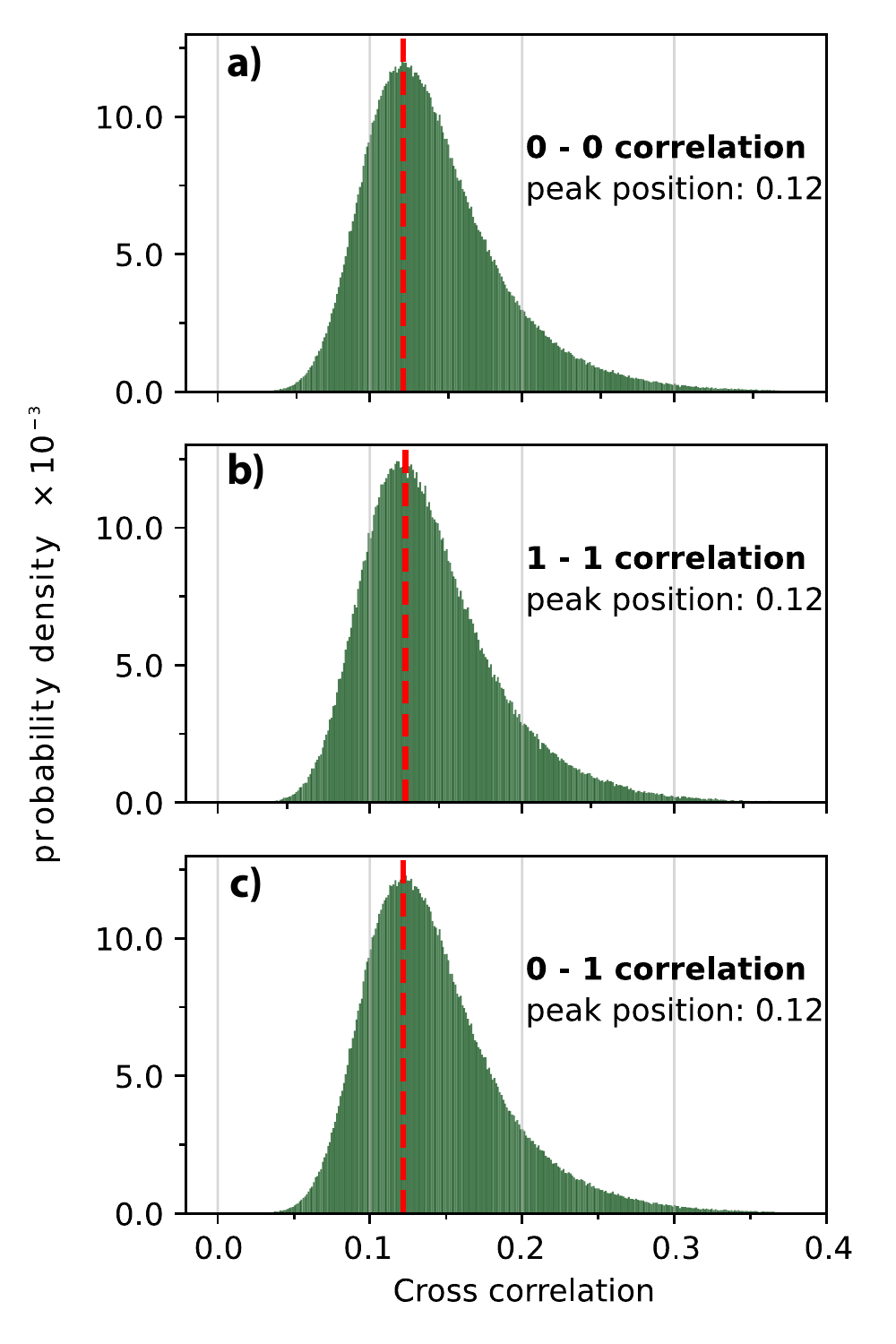}
    \caption{Probability density histograms of speckle cross-correlation between different signals: a) between all pairs of speckle patterns belonging to state 0, b) between all pairs of speckle patterns belonging to state 1, and c) between speckle patterns from state 0 and speckle patterns from state 1. 
    The vertical dashed red lines indicate the peaks of the distributions. 
    }
    \label{fig:correlation}
\end{figure}

When we calculate $C_{\rm P}$ between \textit{all} $N_{\rm W}$ available intermediate speckles, \textit{i.e.}, $C_{\rm P}$ between individual pairs of speckles with all the combinations possible, we get a distribution of $C_{\rm P}$.
In Fig.~\ref{fig:correlation} we show the correlation distribution for the three cases previously described, 0-0, 1-1, and 0-1.
We see that, qualitatively, the distributions are very similar, with the same peak position.

\begin{table}[tbp]
\centering
\caption{K-S statistic between the different correlation distributions.
Each K-S statistic is calculated between Dataset 1 and Dataset 2.   
The $p$-value for every case is on the order of $10^{-15}$.}
\label{tab:KScorr}
\begin{tabular}{l|l|ll}
\textbf{Dataset 1} & \textbf{Dataset 2} & \textbf{K-S statistic} & \textbf{$p$-value} \\ \hline
0-0 correlation    & 1-1 correlation    & 0.03                   & 0.00               \\
0-1 correlation    & 0-0 correlation    & 0.03                   & 0.00               \\
1-1 correlation    & 0-1 correlation    & 0.02                   & 0.00              
\end{tabular}
\end{table}

To compare these distributions quantitatively, we used the two-sample \seqsplit{Kolmogorov}–\seqsplit{Smirnov} (K-S) test~\cite{Kanji2006book}.
This test compares the empirical distribution of two sets of observations. 
When the two observations are from the same distribution, the K-S statistic tends to zero.
The results of this test are shown in Table~\ref{tab:KScorr}. 
We see that the K-S statistic is close to zero and that the $p$-value is lower than 5$\%$ ($\mathcal{O}(10^{-15})$), this tells us that the correlation distributions of the three groups are indistinguishable. 

We see that the correlation distributions are not centered around zero. 
When we claim there is no clear correlation between speckles from the same group, it is because this correlation is the same as the correlation between speckles from different groups. 
However, the fact that the correlation is not zero leads to the question if there is any underlying relation between speckle patterns that the Pearson correlation coefficient does not have access to. 
To test this further, we used unsupervised classification algorithms.

%
%
\section{Unsupervised classification of intermediate speckles}\label{sec:class}
%
%

For our analysis of the intermediate speckle correlations, we use two different classification algorithms: The K-means algorithm and the Hierarchical clustering algorithm. 
 We choose these two algorithms based on our familiarity with them, their required input data, and the ability to visually illustrate their classification process, so that we can have a better understanding of the results.
We use the open-access Scikit-learn Python library to implement both algorithms. 
The procedure for classification is the following: First, to reduce the computational time we reduce the resolution of the picture by averaging the $5 \times 5$ adjacent pixels. 
This is much smaller than the speckle size, thus no information is lost. 
Then, we transform the representation of the data: we consider each pixel as a different dimension and the intensity of this pixel as the position of the realization in that dimension. 
This forms a new high-dimensional space, called the \textit{feature space}. 
Due to the resolution of our camera ($1280\times960$) and the reduced resolution, the feature space has almost $50,000$ dimensions, where each point is a realization (or speckle). 

To reduce even further the dimensionality of the problem and thus make the problem addressable for a personal computer, we use the principal components analysis (PCA) and only consider the first 100 principal components, explaining $>90\%$ of the variance of the data. 
This new data representation is finally used for the classification algorithm, where each principal component is now a dimension of the feature space, and its value corresponds to the position of that dimension. 
Now, the feature space has $N_{\rm d}=100$ dimensions. 

For both algorithms, we use two different classes, intending to separate state 0 and state 1. 
For the K-Means algorithm, we use 100 initializations and a maximum of $30,000$ iterations. 
For the Hierarchical clustering algorithm, the agglomerate strategy is used.
For the sake of generality, the classification is also run without reducing dimensionality, \textit{i.e.}, without using PCA.
For these cases, the image was reduced by averaging the $10 \times 10$ adjacent pixels, considering the data memory compared with the previous cases. 

We show the results of the classification in Table~\ref{tab:classRes}. 
In total, four classification methods are used, two different algorithms and two different data representations. 

To get statistical information from the classification, each method is repeated 5 times with sub-groups of half of the data. 
Furthermore, for each sub-group, we repeat 5 more times to average over the initial random guess of the algorithms.
Note that we select an equal number of speckle patterns from both groups, $N_{\rm W,0}=N_{\rm W,1}=1774$, so the data is evenly distributed between the two groups.
Table~\ref{tab:classRes} presents the balanced accuracy of the classification algorithms. 
The balanced accuracy is an average between the true positive rate (TPR) and true negative rate (TNR), which is the rate of predicted versus the total amount of positives and negatives, respectively~\cite{Brodersen2010icpr}. 
In our case, positive and negative represent a binary 0 and a binary 1.
We see that the balanced accuracy of all the methods we used is around 50\%.
As this is a binary classification and only two options are possible, a random classifier yields on average the same accuracy of 50\%. 
This shows that the classification is as effective as tossing a coin, thus we obtain no new information from it.
Furthermore, the right column of Table~\ref{tab:classRes} shows the percentage of 0's predicted by the clustering. 
For the last method, all the data points were classified as 0, which also yields 50\% accuracy. 
All the other methods separate the data exactly in half for every sub-group and every repetition.

We thus find that there are no trivial correlations between intermediate speckle patterns, regardless of their corresponding encoded message.
Furthermore, the different unsupervised classification algorithms are not able to find any correlation or separation between the patterns.
This means that when measuring two intermediate speckle patterns, is not possible to know for certain if they are encoding the same message or not in the final speckle pattern.
We believe that this knowledge is relevant for applications, particularly in the field of optical wireless communication (OWC), to make communication more secure. 
Therefore, in the following section, we described a possible implementation of a communication link based on these findings.

\begin{table}[tbp]
\centering

\caption{Balanced classification accuracy of the unsupervised methods under study. 
}
\label{tab:classRes}
\begin{tabular}{l|lll}
\textbf{Method}      & \textbf{Average (\%)} & \textbf{STD (\%)} & \textbf{0's predicted (\%)} \\ \hline
K-Means              & 51                    & 4.07              & 50                          \\
Hierarchical        & 49                    & 7.58              & 50                          \\
K-means no PCA       & 49                    & 2.52              & 50                          \\
Hierarchical no PCA & 50                    & 0.00              & 100                        
\end{tabular}

\end{table}

%
%
\section{Proposed communication scheme}\label{sec:scheme}
\begin{figure}[tbp]
    \centering
    \includegraphics[width=\columnwidth]{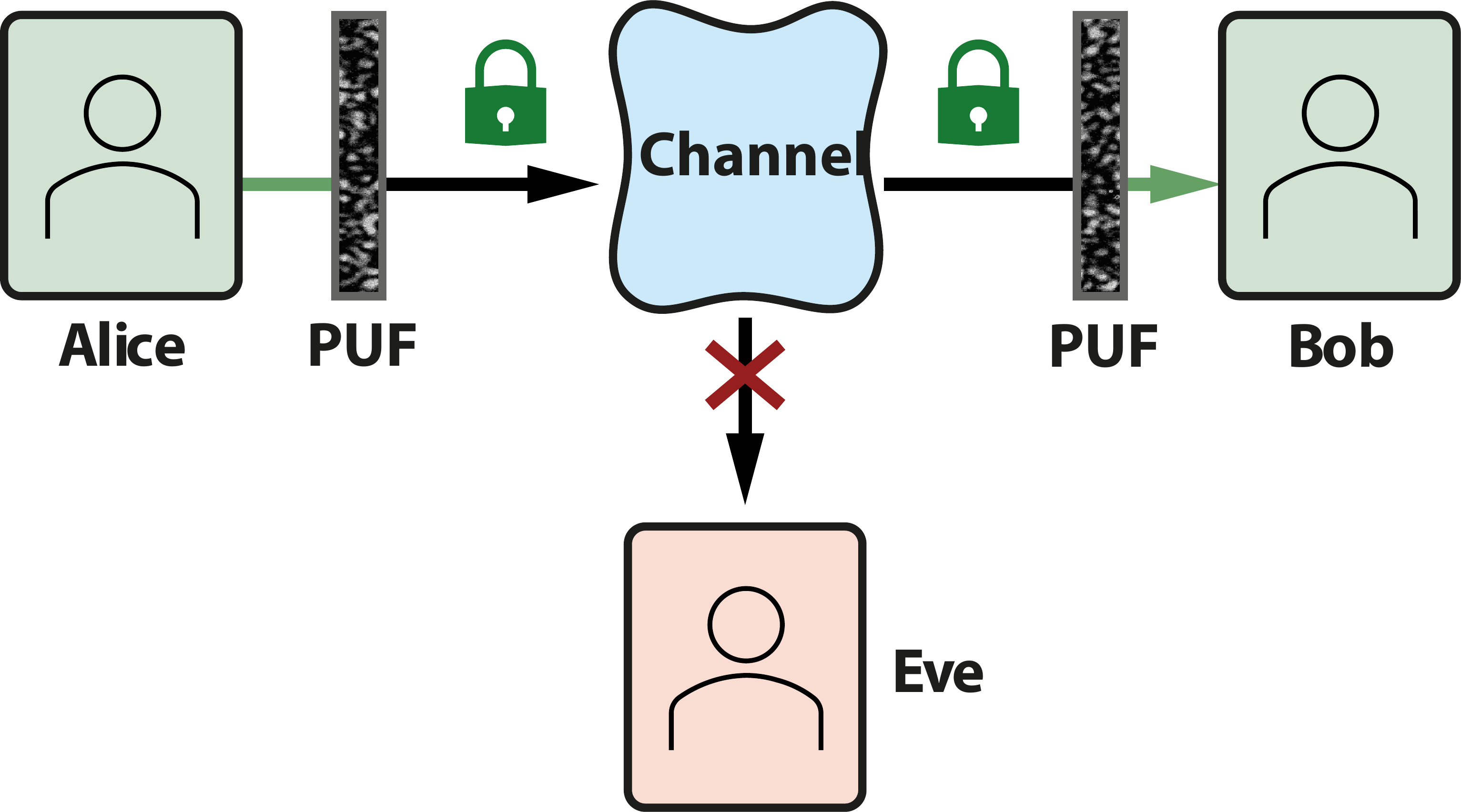}
    \caption{Scheme of the communication link. 
    Alice sends a message to Bob through the channel (free-space), which is encrypted using two Physical Unclonable Functions (PUF). 
    }
    \label{fig:diagram}
\end{figure}

%
%
Based on the described scenario, we propose a new communication scheme based on two layers of physical unclonable functions (PUFs). 
This scheme is depicted in Fig.~\ref{fig:diagram}, which is inspired by the experimental setup shown in Fig.~\ref{fig:setup}. Alice sends a message to Bob through free space using visible or infrared coherent light. 
The initial digital message is encoded as a phase-modulated wavefront that is encrypted by the first PUF. 
When arriving at the destination, the signal passes through a second PUF and the message is recovered as light intensity by Bob. 
If an attacker, depicted in Figure~\ref{fig:diagram} as Eve, intercepts the signal, the message will not be recovered as the second PUF is not known. 
Similarly, if Eve tries to send a false message to Bob, this will not be considered as it does not pass through the first PUF.

The secrecy of the proposed system relies on the lack of correlation between speckle patterns, as demonstrated before. 
If for every binary message, the sender randomly selects one wavefront from a large library with many options, the attacker is not able to separate which speckle patterns are related to a state 0 and which ones are to a state 1. 
The proposed system is also resilient for an attacker to send false messages or store the message for future decryption. 
This is thanks to the first PUF, which scrambles the incoming wavefront. 
If the first PUF is not present, the attacker has access to the incident wavefront and they can measure, store, and replicate it. 
This is not challenging because the spatial complexity of the incident wavefront is limited by the spatial resolution of the modulator. 
Inversely, if the first scattering layer is present, the attacker only has access to the intermediate speckle, which has a much larger spatial complexity and it is highly challenging to record with all its properties and fluctuations~\cite{Pappu2002science}. 
Furthermore, if the attacker wants to send a false message, they will need a modulator device or optical system able to replicate the intermediate speckle pattern with high resolution in order to result in the proper final speckle pattern. 

If the digital codification of the message forces Alice to repeat a specific bit on a predictable basis, \textit{e.g.}, for identification, Eve can record the speckle of these bits to identify each binary 0 and binary 1. 
In this case, the number of messages Eve needs to record to obtain all speckles grows as $\mathcal{O}(N_{\rm W} log(N_{\rm W}))$~\cite{Flajolet1992dam}. 
An implementation without this identification is possible by changing the digital codification to avoid any predictable bit.
Even more, based on the complexity of the scattering material and the degrees of freedom at the wavefront modulation, the number of available wavefronts can be made arbitrarily large, increasing the number of messages needed. 
This also limits Eve to send a false message to Bob.

Many alternatives have been studied to break the secrecy of PUFs employ Machine Learning (ML) techniques, which have proved to be powerful tools for these attacks~\cite{Ruhrmair2010asscomp,Gao2020natelec},  at a large expense of time and digital memory from the side of the attacker.
In most of these studies, however, they needed to obtain a Challenge-Response Pair (CRP) set for training a supervised algorithm. 
To do so, it is assumed that Eve can send a challenge to the PUF and read the response. 
 In our case, the CRP corresponds to the system characterization shown in Section~\ref{sec:charact}.
We obtain all the CRP sets offline and prior to the communication link, as it is end-to-end encryption.
Obtaining a CRP set means that Eve has a subset of speckle patterns at CCD2 with their respective classification at CCD1. 
In our system, an attacker does not have access to the space between the second PUF and CCD1, and gaining access for an invasive measurement poses a threat larger than eavesdropping, thus we disregard such attacks in this study.
If Eve does not have a CRP set, there are no training data for a supervised algorithm. 
Therefore, we do not consider the use of supervised classification methods to be a realistic scenario. 

Nevertheless, an attacker may try to classify the speckle patterns using unsupervised algorithms. 
The results from Section~\ref{sec:class} show that none of the tested unsupervised methods are significantly better than a random classification, which suggests that the data are not classifiable given the intensity speckle pattern obtained from the channel. 
Hence, the communication is secure against an attacker in free-space.

%
%
\section{Applications and limitations}\label{sec:application}
%
We initially envision our scheme in an indoor communication scenario as a one-way communication. 
A central device would include the modulator device with the first PUF layer, and the receiving device would have a pair of photodiodes representing the regions of interest ROI A and ROI B, with its own integrated PUF. 
In this system, the central device can send many messages to different devices using different wavefronts, while the receiver is an inexpensive and fast device. 
The calibration can be done for specific positions of the receiver, \textit{e. g.}, cubicles in an open office. 
If the layout is not changed, the calibration only needs to occur once per device.
We believe our method opens many possibilities to both test applicability of the scheme, and study the scattering properties of complex media. 

Several questions and limitations arise from our system, which need to be addressed to scale up this scheme. 
One limitation of our scheme is the bit rate. 
If only two ROIs are considered, the receiver can consist of two fast photodiodes, thus not limiting the speed. 
The bottleneck arises from the refresh rate of the DMD, which for our DMD model is around 20 kHz. 
This limitation can be remedied by sending multiple bits in one wavefront (\textit{e.g.}, sending 50,000 bits per image results in 1Gbit/s). 
In this scenario, the camera at the receiver is needed, which will limit the speed as well. 
Another solution for higher speed is using another faster modulation technique, like Free-Space Electro-Optic Modulators, which can get up to 100 MHz~\cite{thorlabs2023website}.

Another limitation of our system is the source. 
The need to have a coherent light source hinders us to extend our technique to applications such as light fidelity (Li-Fi). 
This restriction is present in any technology that wants to take advantage of luminaries already installed. 
Nevertheless, it has been demonstrated that modulation techniques such as WFS are applicable to LED sources~\cite{Shao2015spie}. 
Extending our method to LED sources would simplify the architecture. 
Furthermore, it was recently demonstrated that is possible to use a random material as a PUF having a screen projector as a source~\cite{Zhai2023JMO}, which is a commercial combination between an ultra-high-performance (UHP) lamp, which is incoherent, and a DMD. 

Since our system is based on light scattering, it may be highly sensitive to noise, misalignment, or additional scattering events. 
This sensitivity depends on the power of the signal, the distance, and the dynamics of the medium. 
Importantly, we use a diffuser with around 80\% of transmission and a collimating lens, so the use of PUFs does not drastically increase the losses of the system.
This suggests that the restrictions on noise and power are similar to any OWC scheme. 
Furthermore, if the PUF is static and reliable, the dynamics of the medium may fall under the memory effect of the system.
It remains to be tested to what extent these factors limit the applicability of the scheme. 

In case the medium changes significantly or we want to extend this to moving users, the calibration needs to be done online. 
One way of implementing this is adding a secondary classical channel where the users can share the CRP. 
More studies are needed to test the secrecy of such implementation.

Finally, the main advantage of our system is that the secrecy is imposed by the number of measurements needed from an attacker to learn the message. 
This is more restrictive than only computational power, as traditional digital encryption. 
While the results presented in this manuscript indicate no correlation between the signal retrieved by the attacker and the final message, it is important to acknowledge that it is impossible to account for all possible scenarios that could potentially benefit the attacker. 

For future tests, we conceive that an attacker may benefit from other detection configurations that may give more information. 
We think of at least three possible extensions to the proposed detection configuration: (a) measure the complex information of the speckle (both amplitude and phase), (b) measure in the conjugated plane of the sample, and (c) measure the reflective speckle from the receiver PUF. 
The reflective speckle pattern is particularly relevant, as it has been successfully used in imaging through scattering media~\cite{Badon2020science,Lambert2020prx,Lambert2020procnas}. 
It has been proven that the reflective speckle pattern has a certain correlation with the transmitted speckle pattern~\cite{Shahjahan2014apl}, similar to the correlations between the incident and transmitted wavefront mentioned before and may prove useful to decipher the message in our communication scheme. 
Because we calibrate the system beforehand, if the commented configurations give useful information to the attacker (\textit{e.g.}, the reflection-transmission correlations), we believe it is possible to account for it beforehand by filtering further which available wavefronts are available to send the message. 
Furthermore, from the practical point of view, one attractive part of the present scheme is that the detection of the message by the receiver can be done easily, fast, and cheaply. 
This is because the receiver only needs two photodiodes to measure the average intensity. 
In turn, the commented possible configurations impose high technical difficulties. 
Further studies are needed to evaluate how obtaining this new information may affect the secrecy of the scheme. 

%
\section{Conclusion}\label{sec:con}
%
In this paper, we have studied the correlation between speckle patterns when passing through multiple slabs of scattering media. 
We spatially modulate the phase of the incoming light and we send a signal through two diffusers, measuring the resulting speckle pattern both between and after the two diffusers. 
The signal is encoded as changes in light intensity at two regions of interest (ROI) at the receiver, where multiple modulated incoming wavefronts may result in the same message. 
We have studied the correlation between speckle patterns when sending different messages. 
Therefore, we use the Pearson correlation coefficient and two unsupervised classification algorithms. 
In all cases, we observe that there is no correlation between the intermediate speckle pattern and the resulting pattern (or message). 
This method is attractive for optical wireless communication (OWC) schemes, particularly in line-of-sight communication and wireless indoor communication. 

\begin{backmatter}
\bmsection{Funding}
Nederlandse Organisatie voor Wetenschappelijk Onderzoek (NWO-TTW P15-36).

\bmsection{Acknowledgments}
We thank Cornelis Harteveld for expert technical support and sample preparation, and Ad Lagendijk, Allard Mosk, Chris Toebes, and Pepijn Pinkse for stimulating discussions. 
This work was supported by the MESA+ Institute section Applied Nanophotonics (ANP).

\bmsection{Disclosures}
The authors declare no conflicts of interest. 

\bmsection{Data availability} 
The data used for this publication are publicly available in the Zenodo database~\cite{zenodoDatabase}. 
\end{backmatter}

%
%
\bibliography{References.bib}

\end{document}